\documentclass[iop]{emulateapj}
\usepackage{graphicx}
\usepackage{txfonts}
\usepackage{natbib}
\bibpunct{(}{)}{;}{a}{}{,}

\begin{document}

\title{The dark side of penumbral microjets: Observations in H$\alpha$}

\author{D. Buehler}
\affil{Institute for Solar Physics, Dept. of Astronomy, Stockholm University\\ 
Albanova University Center, SE-10961\\ 
Stockholm, Sweden}

\author{S. Esteban Pozuelo}
\affil{Institute for Solar Physics, Dept. of Astronomy, Stockholm University\\ 
Albanova University Center, SE-10961\\ 
Stockholm, Sweden}

\author{J. de la Cruz Rodriguez}
\affil{Institute for Solar Physics, Dept. of Astronomy, Stockholm University\\ 
Albanova University Center, SE-10961\\ 
Stockholm, Sweden}

\author{G. B. Scharmer}
\affil{Institute for Solar Physics, Dept. of Astronomy, Stockholm University\\ 
Albanova University Center, SE-10961\\ 
Stockholm, Sweden}



\date{\today}

\begin{abstract}

We present data of 10 penumbral microjets (PMJs) observed in H$\alpha$, Ca II 8542 \AA, and Fe I 6302 \AA$ $ line pair with the Swedish 1 m Solar Telescope (SST) with CRISP and Ca II K with SST/CHROMIS in active region NOAA 12599 on the $12^{th}$ October 2016 at $\mu=0.68$. All four Stokes parameters of the Ca II 8542 \AA$ $ and Fe I 6302 \AA$ $ lines were observed and a series of test pixels was inverted using the Stockholm inversion code. Our analysis revealed for the first time that PMJs are visible in H$\alpha$, where they appear as dark features with average line-of-sight (LOS) upflows of $1.1\pm0.6$ km/s, matching the LOS velocities from the inversions. Based on the H$\alpha$ observations we extend the previous average length and lifetime of PMJs to $2815\pm530$ km and $163\pm25$ s, respectively. The plane-of-sky (POS) velocities of our PMJs of up to $17$ km/s tend to give increased  velocities with distance travelled. Furthermore, two of our PMJs with significant Stokes $V$ signal indicate that the PMJs possess an increased LOS magnetic field of up to 100 G compared to the local pre-/post- PMJ magnetic field, which propagates as quickly as the PMJs' POS velocities. Finally, we present evidence that PMJs display an on average 1 minute gradual precursory brightening that only manifests itself in the cores of the Ca II lines. We conclude that PMJs are not ordinary jets but likely are manifestations of heat fronts that propagate at the local Alfv\'en velocity. 

\end{abstract}



\section{Introduction} \label{sec:intro}
 
Penumbral microjets (PMJs) are chromospheric brightenings within the penumbral area of a sunspot first described by \citet{katsukawa2007} using Ca II H broadband images observed by the Hinode satellite \citep{kosugi2007}. The brightenings take on an elongated jet-like appearance, and in conjunction with subsequent observations obtained with the Dunn Solar Telescope and the Swedish 1 m Solar Telescope (SST) \citep{scharmer2003}, display distinct emission peaks in the wings of Ca II 8542 \AA$ $ \citep{reardon2013} and enhanced $K_2$ peaks in Ca II K  \citep{esteban2019}. The average length and lifetime of PMJs are 640 km and 90 s, respectively \citep{drews2017}.

Following their discovery, investigatory attention soon focused on the manifestations of PMJs in other spectral lines and parts of the solar atmosphere, beginning with \citet{jurcak2008}, who, using a set of Hinode observations, reported that PMJs propagate along a sunspot's chromospheric magnetic field lines. \citet{katsukawa2010} showed that some photospheric penumbral downflows appear cospatially with PMJs, whereas \citet{esteban2019}, using SST observations, found no distinct PMJ related change in the photosphere compared to its pre-/post- PMJ state. \citet{tiwari2016}, using a set of Hinode, Hi-C \citep{cirtain2013}, and SDO \citep{lemen2012} observations, and \citet{esteban2019}, revealed that PMJs occur preferentially above the photospheric penumbral spine/inter-spine boundary \citep[see e.g.][]{lites1993,solanki2003rev}. \citet{vissers2015} and \citet{samanta2017}, both employing IRIS \citep{depontieu2014} observations, demonstrated that PMJs produce significant heating in the solar transition region, but manifestations of PMJs in the solar corona have so far not been made \citep{tiwari2016}. Furthermore, \citet{tiwari2016}, \citet{drews2017}, and \citet{esteban2019} revealed that PMJ locations are not homogeneously distributed across the penumbra but tend to be clustered. 

Given the jet-like appearance of PMJs, measurements of their velocities form a constituent part of their analysis. However, discrepancies between their line-of-sight (LOS) and plane-of-sky (POS) velocities soon became apparent. The reported LOS velocities range between  $\pm4$ km/s \citep{esteban2019} and are generally slower \citep{reardon2013,vissers2015,drews2017} than the POS velocities that range up to $100$ km/s \citep{katsukawa2007}.

The rapid brightening, short lifetime, and elongated appearance of PMJs have driven their physical interpretation in terms of a reconnection scenario. \citet{katsukawa2007} proposed reconnection between neighbouring vertical and horizontal magnetic fields in the penumbra, which produces heating and drives a jet into the chromosphere. \citet{tiwari2016} modified this scenario by suggesting the reconnection takes place at the spine/inter-spine boundary based on observations of photospheric return flows \citep{zakharov2008,joshi2011,scharmer2011,tiwari2013} and opposite polarity magnetic fields in the penumbral inter-spines \citep{ruizcobo2013,scharmer2013} and advances in the simulated \citep{sakai2008,magara2010} magnetic structure of the penumbra. \citet{ryutova2008} suggested that a PMJ is a manifestation of a flux tube that rises upward supersonically under magnetic tension following a reconnection event in the penumbra. \citet{samanta2017} moved the location of the reconnection into the solar transition region based on IRIS observations that revealed a bright point at the PMJ location before the PMJ occurs, which was absent in cospatial Hinode broadband observations. \citet{esteban2019} proposed, in an effort to reconcile the wide range of LOS and POS velocity measurements and the complex LOS velocity stratifications, that following a reconnection in the photosphere, a PMJ is a rapidly propagating perturbation front akin to the heat fronts present by \citet{depontieu2017} based on a simulation of type II spicules \citep{martinezsykora2017}. 

However, unequivocal evidence in support of a reconnection scenario is lacking so far, owing in large part to the exceeding observational difficulties associated with the reconnection scenario. A first step may have been taken in the observations by \citet{tiwari2016,tiwari2018}, who described the presence of opposite polarity patches below their 'large' PMJs and tentative evidence of a reduction in magnetic flux following the PMJ.

In this investigation, featuring a range of spectral lines simultaneously observed with the SST, we present the previously unreported manifestation of PMJs in the Balmer H$\alpha$ line, which led us to an extension and reappraisal of PMJ properties that are also presented in the paper. 
 
\section{Observations and Analysis} \label{sec:obs}

Active region NOAA 12599 was observed with the SST using the CRISP \citep{scharmer2008} and CHROMIS Fabry-Perot imagers. At the time of observation on the $12^{th}$ October 2016 the active region was located near $630''$ X and $-320''$ Y with $\mu=0.68$ and observed continuously for 30 minutes beginning at 10:57:58 UT. 

CRISP sequentially recorded the H$\alpha$, Ca II at 8542 \AA, and Fe I at 6301 and 6302 \AA$ $ absorption lines. The H$\alpha$ line was sampled at 17 wavelength points in steps of 150 m\AA$ $ between $\pm$1050 \AA$ $ around the line core along with two far wing positions at $\pm$1550 \AA. The 6302 Fe I line was sampled between $\pm$120 m\AA$ $ in steps of 40 m\AA$ $ along with two wing positions at $\pm$190 m\AA$ $ around the line core. The 6302 Fe I line was only sampled in steps 40 m\AA$ $ within $\pm$80 m\AA$ $ around the line core. Additionally, a single continuum point was sampled between the two lines, forming a total of 16 wavelength points. The Ca II 8542 \AA$ $ was sampled between $\pm$765 \AA$ $ in steps of 85 m\AA$ $ along with two wing positions at $\pm$1550 m\AA$ $ amounting to 21 wavelength points. The H$\alpha$ line was only sampled in Stokes $I$, whereas all four Stokes parameters, $I$, $Q$, $U$, and $V$, were obtained for the Fe I and Ca II lines, leading to a total acquisition time of 39 s for the entire CRISP sequence. 

CHROMIS recorded the Ca II K and H$\beta$ absorption lines. The Ca II K line at 3934 \AA$ $ was sampled between $\pm$1020 m\AA$ $ in steps of 60 m\AA$ $ in addition to a continuum point at 4000 \AA$ $ amounting to 36 wavelength points in total. The H$\beta$ data were not used in this investigation. The entire CHROMIS sequence required 13 s, leading to three CHROMIS scans for every CRISP scan. 

Each data set was reduced using the CRISPRED pipeline \citep{delacruz2015}, which includes the Multi-Frame Blind Deconvolution (MOMFBD) image restoration technique \citep{vannoort2005,vannoort2008}. The CRISPRED pipeline was extended for the CHROMIS data and additionally included a phase-diversity reconstruction step \citep{lofdahl2018}.  Finally, all the data sets were aligned and de-rotated. The CRISPEX analysis tool \citep{vissers2012} was used extensively during our investigation.

During our analysis we performed a series of test inversions to gain a more thorough understanding of the solar atmosphere. The employed inversion code, STiC \citep{delacruz2016,delacruz2019}, is based on the RH synthesis code \citep{uitenbroek2001} which is capable of dealing with non-LTE environments including the partial redistribution of scattered photons. The code employs a Levenberg-Marquardt algorithm to iteratively minimise the $\chi^2$ between the observed and synthesised spectra and the resulting output atmospheres are in hydrostatic equilibrium.

We simultaneously inverted all four Stokes parameters of the Fe I and Ca II 8542 \AA$ $ lines. Initially, we performed several trials using different node setups in $\log(\tau)$ to determine a node placement that prevented the code from retrieving unphysical solutions whilst allowing it to fit the numerous asymmetries present in the observed spectra resulting from the strong gradients present in the solar atmosphere. The final $\log(\tau)$ node setup was as follows (where $\tau$ refers to the continuum optical depth at 5000 \AA): temperature: [-5.5,-5,-4,-2,-0.8,0]; microturbulence: [-3,-2,-0.8,-0.2]; LOS velocity: [-5.5,-5,-4,-2,-0.2]; LOS magnetic field: [-4,-2]; POS magnetic field: [-4,-2]. All inversions were repeated three times with the same node setup, but using a slightly randomised atmosphere obtained from the pixel's previous inversion to prevent the minimisation algorithm from falling into a local minimum.  

\section{Results} \label{sec:results}

\begin{figure}
\includegraphics[width=8cm]{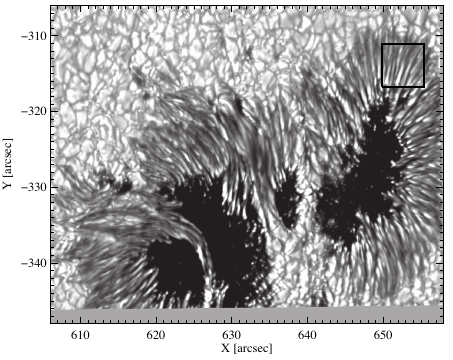}
\includegraphics[width=8cm]{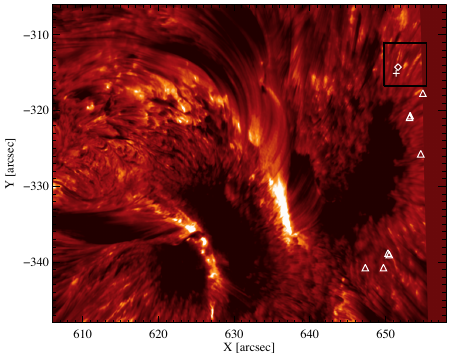}
\caption{Overview of the sunspot in active region NOAA 12599 as it appears in the blue continuum and in H$\alpha$ -450 m\AA. The black square and plus symbol in the H$\alpha$ image indicate the size and location of the FOV in Fig. \ref{fig:composite2}. The diamond symbol refers to the PMJ displayed in Fig. \ref{fig:compositeappen}.  The white triangle symbols mark the locations of the remaining analysed PMJs. \label{fig:composite1}}
\end{figure}

Active region NOAA 12599 had already emerged on the solar surface when it first appeared on the East limb and new flux emerged intermittently during its passage across the solar disc. By the time the active region was observed with the SST a fully developed sunspot had formed (see continuum image in Fig. \ref{fig:composite1}) and its penumbra and light bridges continued to evolve markedly until the active region disappeared behind the west limb. The chromosphere above the sunspot featured copious amounts of jetting, in particular above the light bridges, which obscured parts of the surrounding penumbra (see the H$\alpha$ image in Fig. \ref{fig:composite1}). The limbward facing penumbra of the sunspot was least affected by the jetting and hence best suited to clearly identify PMJs and their respective manifestations in the observed lines. We identified 10 PMJs in total based on their distinct spectral shapes in Ca II K and Ca II 8542 \AA, and their locations are marked by the triangles in Fig. \ref{fig:composite1}. 

\begin{figure*}
\includegraphics[width=18cm]{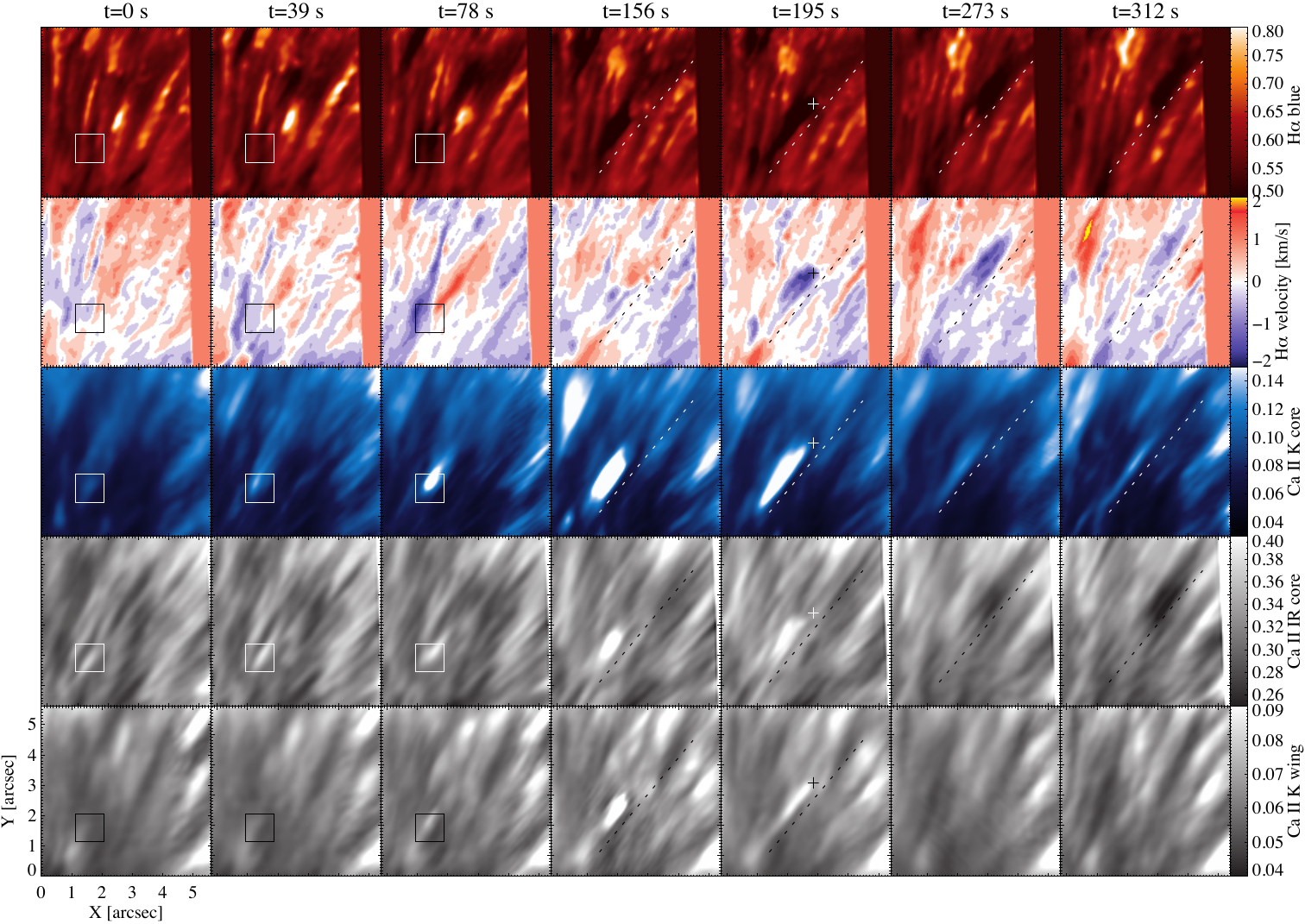}
\caption{Sequence of H$\alpha$ -450 m\AA, H$\alpha$ bisector LOS velocities at -450 m\AA, Ca II K -60 m\AA, Ca II 8542 -85 m\AA, and Ca II K -300 m\AA$ $ images. The squares enclose the origin of the PMJ that occurs at $t=78$ s, and along with the plus symbol mark the pixels that have been inverted. The dotted lines are in parallel with the PMJ's propagation axis. The animation starts at $t=0$ s and ends at $t=390$ s. \label{fig:composite2}}
\end{figure*}

In the following we will focus on the PMJ marked by the plus symbol and enclosed by the black square in Fig. \ref{fig:composite1} as an illustrative example, but the general properties and behaviours of this PMJ apply to all the examples we identified. The appendix features an additional PMJ example. The sequence of images displayed in Fig. \ref{fig:composite2} tracks the temporal evolution of the PMJ at different wavelengths. The PMJ occurs at $t=78$ s and spawns near the spine/inter-spine boundary, which is most clearly visible in the H$\alpha$ panels in Fig. \ref{fig:composite2}. It displays the familiar elongated brightening in both the Ca II K and Ca II 8542 \AA$ $ lines, which lasts less than 2 minutes in the wings of both lines. Unlike previous observations we also find a cospatial and cotemporal response to the PMJ in H$\alpha$ in the form of a dark elongated feature. The dark feature in H$\alpha$ continues to propagate after the brightening in the wings of the Ca II lines has already subsided. It eventually becomes visible in the H$\alpha$ line core at $t=312$ s (not displayed in Fig. \ref{fig:composite2}) indicating that the PMJ also has a LOS velocity component in addition to the POS component displayed in Fig. \ref{fig:composite2}. The PMJ always appears as a 'dark' feature in H$\alpha$ and in any H$\alpha$ wavelength position it can be identified, at all times. Movie 1 also shows the evolution of the jet in Fig. \ref{fig:composite2}.

A closer inspection of the Ca II images in Fig. \ref{fig:composite2} reveals that the PMJ continues to remain visible in the cores of both Ca II lines beyond the initial brightening in line with H$\alpha$ in the form of a dark feature. However, the apparent darkness of the PMJ in the Ca II lines cores is caused by a blue-shift of both line cores. The PMJ is a 'bright' feature relative to its immediate surroundings throughout its lifetime, when traced in the red at, e.g. Ca II 8542 \AA + 85 m\AA. Parts of the PMJ always appear dark in any Ca II K wavelength due to the complex interplay between the $K_2$ and $K_3$ intensity peaks and their respective Doppler velocity shifts.

The comparatively longer lifetime of the PMJ in the cores relative to the wings of the Ca lines is another indication that the PMJ moves up as well as outward during its lifetime. The PMJ is untraceable in either the blue or red wings of both Ca II lines following the initial brightening, as Fig. \ref{fig:composite2} suggests.

In addition to the intensity images, Fig. \ref{fig:composite2} also displays LOS velocities obtained from bisectors of the H$\alpha$ profiles. The bisectors were calculated from spline interpolated profiles and the velocity calibration was based on a 30 minute averaged umbral H$\alpha$ profile obtained from the nearby umbra in the observations. The PMJ reveals itself as a feature hosting upflows of up to 2 km/s in H$\alpha$. The upflows in Fig. \ref{fig:composite2} are consistently higher at the edges of the PMJ compared to its centre and indicate that this PMJ has some internal structure that is also visible in the intensity images (e.g. Ca II 8542 at $t=273$ s in Fig. \ref{fig:composite2}). However, not all the PMJs we identified possessed such apparent internal structure. Nonetheless, all of the 10 PMJs that we analysed display upflows in H$\alpha$ and the average LOS velocity of all PMJs is $-1.1\pm0.6$ km/s.

We also proceeded to measure the lifetime and maximum distance travelled by the PMJ in Fig. \ref{fig:composite2}, which are $3547$ km and $234$ s, respectively. The average lifetime and maximum distance averaged over all 10 PMJs amounts to $2815\pm530$ km and $163\pm25$ s, respectively. Table \ref{tab:properties} in the appendix lists the individual properties of all 10 PMJs. Given that many of our selected PMJs are close to the edge of the field of view (FOV), as Fig. \ref{fig:composite1} indicates, it is possible that our averaged values underestimate the true average lifetime and maximum distance travelled by our PMJs.

\begin{figure}
\includegraphics[width=8cm]{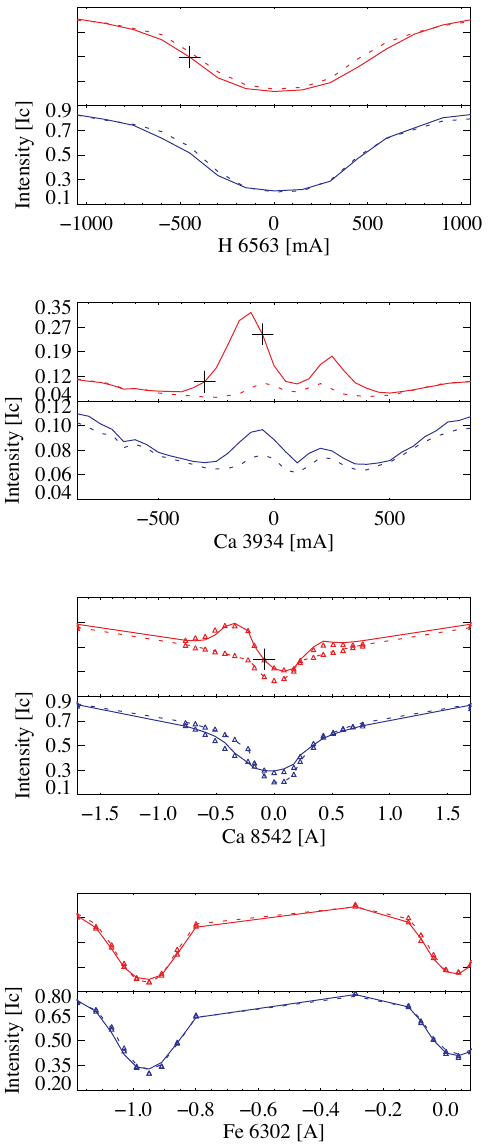}
\caption{ Series of H$\alpha$, Ca II K, Ca II 8542 \AA, and Fe I 6302 \AA $ $ spectra taken at the origin (red spectra) and plus symbol (blue spectra) of the PMJ in Fig. \ref{fig:composite2}. The dotted lines are spectra taken before the PMJ occurs at $t=0$ s and the solid lines are spectra taken during the PMJ at $t=78$ s and $t=216$ s for the red and blue spectra, respectively, following the timeline in Fig. \ref{fig:composite2}. The dotted and solid lines in the Ca II 8542 \AA$ $ and Fe I 6302 \AA$ $ panels are the fits obtained from the STiC inversions and the triangles indicate the measured intensities. All the spectra were normalised to their respective the quiet-Sun continuum values. The large plus symbols indicate the respective spectral positions of the images displayed in Fig. \ref{fig:composite2}. \label{fig:spectra}}
\end{figure}

The red spectra shown in Fig. \ref{fig:spectra} are taken at the origin pixel of the PMJ displayed in Fig. \ref{fig:composite2} and the dotted lines display the spectra at $t=0$ s before the PMJ occurs and the solid lines in each panel display the spectra at $t=78$ s when the PMJ occurs. The panels of both Ca II lines and the Fe I line illustrate the spectral characteristics common to all PMJs, an intensity reversal in the wings of Ca II 8542 \AA, a marked intensity increase and broadening of the $K_2$ peaks in the Ca II K line, and an insignificant change in the photospheric Fe I lines. Both Ca II lines also feature a blue over red asymmetry in their line profiles, which is common for the 10 PMJs that we selected but not to all PMJs in general \citep[see][]{drews2017}.

At the PMJ origin pixel, located at the centre of the box in Fig. \ref{fig:composite2}, the red solid H$\alpha$ profile at $t=78$ s shown in Fig. \ref{fig:spectra} is broader compared to the red dotted pre-PMJ profile also displayed in the same panel. At the later stages of the PMJ's evolution the blue solid H$\alpha$ profile in Fig. \ref{fig:spectra} hosts a blue-shift compared to its local blue dotted pre-PMJ profiles, as suggested by the bisector velocities in Fig. \ref{fig:composite2}. The blue solid Ca II 8542 \AA$ $ spectra in Fig. \ref{fig:spectra} at $t=234$ s taken at the plus symbol in Fig. \ref{fig:composite2} feature a pure absorption profile whose core is shallower and blue-shifted compared to its local pre-PMJ profile; see the blue dotted line in the same panel. The blue solid Ca II K line at $t=216$ s displays reduced $K_2$ peak intensities compared to the red solid line profile in Fig. \ref{fig:spectra}, but both $K_2$ and $K_3$ peaks are brighter when compared to their respective local pre-PMJ profile, see dotted Ca II K lines in the same panels. 

\begin{figure}
\includegraphics[width=8cm]{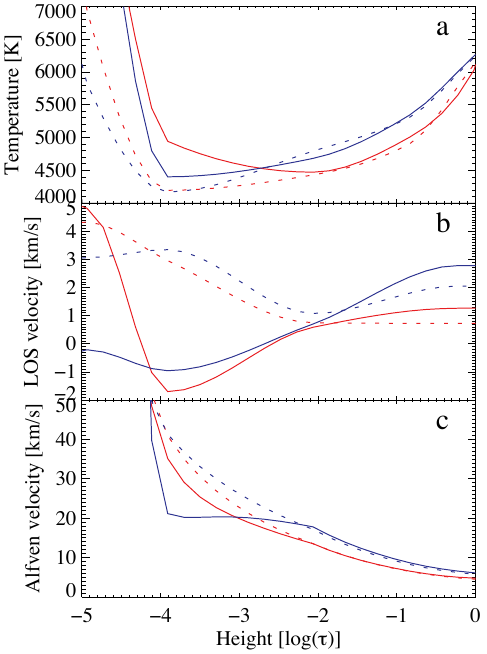}
\caption{Stratifications of the temperature (a), LOS velocity (b), and Alfv\'en velocity (c), from the STiC inversions obtained from the fits displayed in Fig. \ref{fig:spectra}. The red dotted and solid temperature stratifications belong to the origin pixel of the PMJ in Fig. \ref{fig:composite2} at $t=0$ s and $t=78$ s, respectively. The blue dotted and solid temperature stratifications belong to the pixel at the plus symbol in Fig. \ref{fig:composite2} at $t=0$ s and $t=216$ s, respectively. \label{fig:strat}}
\end{figure}

The dotted and solid lines in the Ca II 8542 \AA$ $ and Fe I 6302 \AA$ $ panels in Fig. \ref{fig:spectra} are the fits to the measured spectra, shown by the triangles in the same panels, produced by STiC. The temperature stratifications pertaining to these fits are displayed in Fig. \ref{fig:strat}a. The solid red line in the figure demonstrates the temperature increase seen in the chromosphere during the PMJ compared to the dotted red line, which shows the pre-PMJ temperature stratification at the PMJ origin. The solid blue stratification indicates that the PMJ still causes a temperature increase in the chromosphere by the time it reaches the position corresponding to the plus symbol in Fig. \ref{fig:composite2} compared to the local pre-PMJ temperature stratification shown in by the blue dotted line; however, the heating is weaker compared to the red solid line within the $\log(\tau)=-2$ to $-4$ range. Also, the blue solid line temperature stratification in Fig. \ref{fig:strat}a only begins to distinguish itself from its dotted counterpart above $\log(\tau)=-3$ compared to $\log(\tau)=-2$ for the red pair, which supports the conclusion drawn from Fig. \ref{fig:composite2} that the PMJ appears to move up as it propagates. 

The H$\alpha$ bisector LOS velocities in Fig. \ref{fig:composite2} associate the displayed PMJ with upflows ranging between $1$ and $2$ km/s during its lifetime. The PMJ LOS velocities from the inversions of the pixels in Fig. \ref{fig:spectra}, displayed in Fig. \ref{fig:strat}b by the solid lines, support the inferred H$\alpha$ velocities. The red solid velocity stratification shows upflows below $\log(\tau)=-4$ but downflows above $\log(\tau)=-4$, which support the H$\alpha$ observations that also display downflows at $t=78$ s in Fig. \ref{fig:composite2} at the origin pixel. The majority of our selected PMJs host downflows in H$\alpha$ at their origin before the PMJ occurs. The photospheric Evershed flow is responsible for the persistent downflows below $\log(\tau)-2$ displayed by all stratifications in Fig. \ref{fig:strat}b.   

\begin{figure}
\includegraphics[width=8cm]{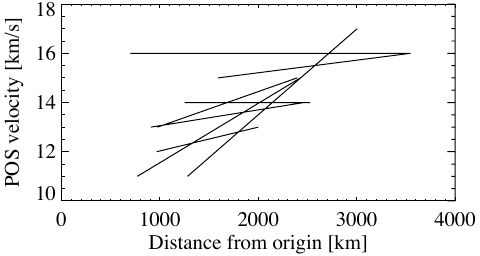}
\caption{POS velocities of the PMJs marked in Fig. \ref{fig:composite1} at various distances from their respective origins. The POS velocity of each PMJ was measured twice. \label{fig:POSvel}}
\end{figure}

\begin{figure}
\includegraphics[width=8cm]{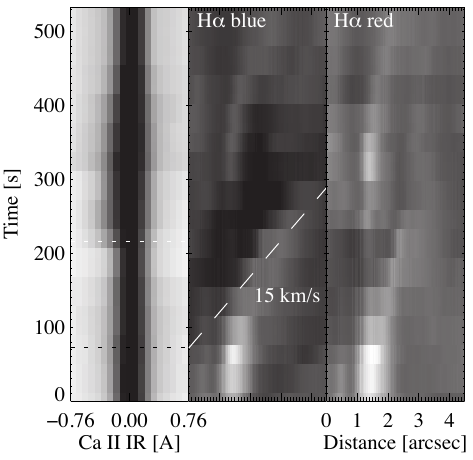}
\caption{Left panel: evolution of the Ca II 8542 \AA$ $ spectrum at the location of the plus symbol in Fig. \ref{fig:composite2}. The dotted black horizontal line marks the start of the PMJ and the white dotted horizontal line marks the time when the PMJ reaches the location of the plus symbol in Fig. \ref{fig:composite2}. Middle and right panels: spacetime diagrams of the PMJ in Fig. \ref{fig:composite2} at H$\alpha$ -300/+450 m\AA. The slice was drawn in parallel with the dotted line in Fig. \ref{fig:composite2} starting at the PMJ origin. The white dashed line indicates a POS velocity of 15 km/s. \label{fig:xt}}
\end{figure}

The PMJ in Fig. \ref{fig:composite2} propagates across the FOV during its lifetime, allowing us to estimate its POS velocity. The comparatively high contrast in H$\alpha$ between the PMJ and the surrounding chromosphere made it the ideal line to track the POS velocity of the PMJ. We measured the POS velocity of the PMJ at $t=195$ s, i.e. near half its lifetime, and again at $t=312$ s, i.e. at the end of its lifetime and the POS velocities are $15$ km/s and $16$ km/s, respectively. We proceeded by measuring the POS velocities of all 10 selected PMJs in the same manner and the resulting POS velocities are displayed in Fig. \ref{fig:POSvel}. The majority of PMJs display an increase of several km/s in their POS velocity during their lifetime and none of them show a decrease in their POS velocity. The measured LOS velocities by comparison are $1-2$ km/s in magnitude and only vary by some $0.5$ km/s throughout the PMJs' lifetimes, indicating that the POS velocity increase shown in Fig. \ref{fig:POSvel} is not at the expense of the measured H$\alpha$ LOS velocities. Furthermore, none of the selected PMJs show a returning downflow either during or after their lifetimes in any of the observed lines. Fig. \ref{fig:xt} demonstrates this behaviour for the PMJ displayed in Fig. \ref{fig:composite2}. Whilst the measured POS velocities are greater than the sound speed in the photosphere of $8$ km/s and therefore supersonic, they are insufficient to overcome solar gravity and should therefore produce return flows. Since no return flows could be detected, the PMJs are unlikely to be ordinary jets typically associated with mass flows.

Therefore, the selected PMJs likely are heat fronts propagating through the chromosphere rather than true jets containing large-mass flows. Such a heat front would propagate at the local Alfv\'en velocity, which we calculated from the inversions of the spectra in Fig. \ref{fig:spectra} and is displayed in Fig. \ref{fig:strat}c. The PMJ in Fig. \ref{fig:composite2} passes through the location corresponding to the plus at $t=234$ at which point its POS velocity would be above $15$ km/s, i.e. $>75 \%$ of the local Alfv\'en velocity of $20$ km/s (see blue solid line between $\log(\tau)= -3$ and $-4$ in Fig. \ref{fig:strat}c). As the POS velocity only measures the horizontal component of the PMJ's propagation, its true propagation velocity would be even higher. Given that the heat front is not associated with large-mass flows the LOS velocities retrieved from the H$\alpha$ bisectors would severely underestimate the true vertical propagation velocity of the PMJ's heat front. An analysis of simultaneously acquired and inverted transition region spectral lines would have helped to constrain the vertical propagation velocity of the PMJ, but were not available at the time of the observation.

\subsection{PMJ magnetic field} \label{sec:mag}

The polarimetric measurements of the Ca II $8542$ \AA$ $ line in principle allow us to retrieve the chromospheric magnetic field in the PMJs. However, the Stokes $Q$ and $U$ amplitudes are never above $3\sigma$ for any of the selected PMJs at any time, meaning that the horizontal magnetic field retrieved by the inversions overestimates the true horizontal magnetic field component and the azimuthal orientation of the magnetic field also remains unknown. Only two PMJs, including the one displayed in Fig. \ref{fig:composite2}, hosted Stokes $V$ signals above $3\sigma$, but only close to their respective origins. 

All the spectra displayed in Fig. \ref{fig:spectra} had Stokes $V$ profiles with amplitudes above $3\sigma$. The inversions of red spectra in Fig. \ref{fig:spectra} retrieve a LOS magnetic field strength that drops from 700 G in the photosphere to 550 G and 450 G in the chromosphere for the solid and dotted lines, respectively. The blue spectra in Fig. \ref{fig:spectra} have LOS magnetic field values of 800 G in the photosphere, which drop to 470 G and 350 G in the chromosphere for the solid and dotted lines, respectively. The total photospheric magnetic field strength is 1 kG and 1.2 kG for all the red and blue spectra in Fig. \ref{fig:spectra}, respectively, and drops to 700 G on average in the chromosphere, but, given the insignificant Stokes $Q$ and $U$ signals in the Ca II 8542 \AA$ $ line, the chromospheric horizontal magnetic field values are potentially too high and unreliable. 

\begin{figure}
\includegraphics[width=8cm]{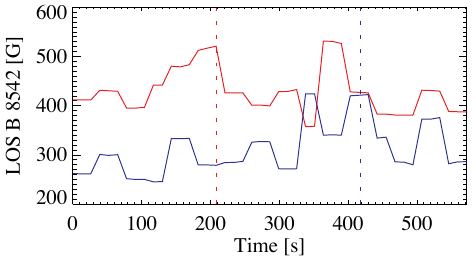}
\caption{Evolution of chromospheric LOS magnetic field, using the weak field approximation on Ca II 8542 \AA, of the PMJ displayed in Fig. \ref{fig:composite2}. The solid red and solid blue lines display the LOS magnetic field evolution at the PMJ origin and the plus symbol, respectively. The dotted vertical lines mark the times when the PMJ spectra in Fig. \ref{fig:spectra} were taken, i.e. at $t=78$ s and $t=216$ s relative to the timeline in Fig. \ref{fig:composite2}. \label{fig:weakB}}
\end{figure}

We proceed by calculating the chromospheric LOS magnetic field from the Ca II 8542 \AA$ $ Stokes $V$ profiles for the FOV displayed in Fig. \ref{fig:composite2} using the weak field approximation \citep[e.g.][]{landi2004,centeno2018}. The resulting magnetic field evolution at the PMJ origin based on a $3\times3$ pixel average around the origin pixel is displayed by the red solid line in Fig. \ref{fig:weakB}. The displayed magnetic field evolution used the entire Stokes $V$ spectrum, but weak field calculations that excluded the Stokes $V$ line core produced qualitatively similar results. The blue solid line in the same figure displays a similar average around the plus symbol seen in Fig. \ref{fig:composite2}. Both lines indicate that the PMJ in Fig. \ref{fig:composite2} can be associated with a 100 G increase in the LOS magnetic field strength compared to the average pre-/post- PMJ field. Furthermore, in conjunction with the two dotted lines in Fig. \ref{fig:weakB}, it becomes apparent that the magnetic field 'bump' travels at a similar POS velocity as the PMJ's intensity signature seen in Fig. \ref{fig:composite2}. The second PMJ in our sample, which also hosted Stokes $V$ amplitudes above $3\sigma$, displays a qualitatively similar LOS magnetic field evolution as shown in Fig. \ref{fig:weakB}, albeit with smaller field strengths.

The lower absolute LOS magnetic field strengths, obtained with the weak field approximation and displayed in Fig. \ref{fig:weakB}, compared to the values retrieved by the inversion, can be explained by the presence of magnetic and velocity gradients in the solar atmosphere, which decrease the amplitude of the Stokes $V$ signal and are only taken into consideration by the inversion but not by the weak field approximation. Furthermore, the Alfv\'en velocity is proportional to the magnetic field, which we most certainly overestimate because of the noisy Stokes $Q$ and $U$ profiles. Therefore, the Alfv\'en velocities above $\log(\tau)=-2$ in Fig. \ref{fig:strat}c are liable to be somewhat overestimated. Also, the fact that our upper magnetic field node in the inversion is placed at $\log(\tau)=-4$, makes the retrieved Alfv\'en velocities above that height even less reliable.

\subsection{PMJ precursor} \label{sec:precursor}

\begin{figure}
\includegraphics[width=8cm]{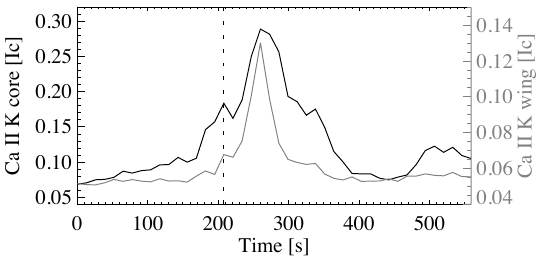}
\caption{Evolution of Ca II K intensity at the origin pixel of the PMJ displayed in Fig. \ref{fig:composite2} at two spectral positions marked by the plus sign in the Ca II K spectra in Fig. \ref{fig:spectra}. The dashed vertical marks the start of the PMJ as defined in Fig. \ref{fig:composite2}. \label{fig:CaI}}
\end{figure}

The Ca II K and 8542 \AA$ $ core images in Fig. \ref{fig:composite2} display a bright point at the PMJ origin before the PMJ occurs at $t=78$ s, which is not visible in either the Ca II K wing observations, also shown in Fig. \ref{fig:composite2}, or in Ca II 8542 \AA$ $ wing observations. The two lines drawn in Fig. \ref{fig:CaI} follow the Ca II K intensity evolution at the PMJ origin from a $3\times3$ pixel average around the origin pixel sampled at the wavelength positions marked in Fig. \ref{fig:composite2} in more detail. The black line illustrates that the Ca II K core begins to gradually brighten relative to the Ca II K wing, shown by the grey line, near $t=60$ s in Fig. \ref{fig:CaI} before the pixel brightens more rapidly in both line positions around $t=170$ s, reaching a simultaneous intensity maximum at $t=260$ s. The initial gradual brightening phase is only detectable near the line core, lasting for 2 minutes for the PMJ in Fig. \ref{fig:composite2} and 1 minute when averaged across all 10 selected PMJs, which all exhibit the phenomenon. Inversions of successive time steps of the origin pixel of the PMJ in Fig. \ref{fig:composite2} reveal that the gradual Ca II K core brightening phase in Fig. \ref{fig:CaI} is associated with a gradually rising temperature above $\log(\tau)=-4$.  

\section{Discussion} \label{sec:dis}

The results presented in this analysis demonstrate that PMJs are detectable in H$\alpha$, but since they appear in absorption at all times, a reliable detection of a PMJ in H$\alpha$ greatly benefits from simultaneous observations in other chromospheric lines such as Ca II H or in our case Ca II K and Ca II 8542 \AA.

The added diagnostics provided by the H$\alpha$ observations allowed us to extend the typically reported PMJ length and lifetime to $2815\pm530$ km and $163\pm25$ s, respectively. Consequently our average PMJ length is longer than the values reported by \citet{vissers2015}, \citet{drews2017}, and \citet{esteban2019}. However, PMJs are typically selected manually, except by \citet{drews2017}, leading to a wide range of lengths of up to 10,000 km \citep{katsukawa2007}, which is longer than any individual PMJ in our sample. The previously reported PMJ lifetimes of $90$ s \citep{katsukawa2007,vissers2015,tiwari2016,drews2017} are nearly doubled by our investigation.

The longer PMJ lifetime and their comparatively high contrast in H$\alpha$ allowed us to not only determine their POS velocity but also to track the change in their POS velocities over time. All 10 selected PMJs revealed either constant or more often increasing POS velocities during their lifetimes (see Fig. \ref{fig:POSvel}). Additionally, we did not observe any return flow for any PMJ in any of the observed spectral lines. Such a POS velocity evolution is uncharacteristic of typical jetting because the measured velocities are far below the solar escape velocity. The acoustically driven jets in the form of dynamic fibrils observed by \citet{depontieu2007df} and simulated by \citet{hansteen2006} and \citet{heggland2011} have similar lengths, lifetimes, and POS velocities similar to the PMJs we observed, in particular when compared to the dynamic fibrils found in sunspots \citep{vandervoort2013}. All the dynamic fibrils follow parabolic paths over their lifetimes, which is not true for our PMJs. Other reconnection driven jets in sunspots \citep{robustini2016} feature far greater velocities than our PMJs but nonetheless always display a return flow. On-disk observations of type II spicules \citep{vandervoort2009} also often do not display return flows, but unlike our PMJs they feature faster upflows in excess of $50$ km/s, which express themselves as doppler shifts in spectral lines such as H$\alpha$ \citep{vandervoort2009,sekse2012,vandervoort2015}. The fact that our PMJs display an average LOS H$\alpha$ velocity upflow of only $1.1\pm0.6$ km/s makes our PMJs not comparable to a typical type II spicule.

However, some spicule type II observations also revealed a discrepancy between their POS \citep{sekse2013,tian2014,narang2016} and LOS velocities \citep{vandervoort2009,sekse2013}. A solution was proposed by \citet{depontieu2017} whereby the comparatively faster POS velocities are a manifestation of independently occurring currents propagating along magnetic field lines at the local Alfv\'en velocity and driven by a combination of magnetic tension and transverse waves. The current dissipates due to ambipolar diffusivity, causing a cospatial heat front. The apparent POS velocity increase of our PMJs is most easily reconciled if the PMJs propagate along the sunspot's chromospheric magnetic field, which becomes more horizontal toward the penumbra/quiet-Sun boundary \citep{joshi2017}, thereby increasing the horizontal velocity component over time at the expense of the vertical velocity component, which is not fully retrieved by our H$\alpha$ LOS velocities. Evidence that PMJs propagate along magnetic field lines was given by \citet{jurcak2008}. The Alfv\'en velocities calculated from our STiC inversions are similar to chromospheric Alfv\'en velocities presented by \citet{depontieu2001} and \citet{grant2018} and further supported the idea that our PMJs likely travel at the local Alfv\'en velocity. Unlike the simulations by \citet{depontieu2017} or X-ray observations by \citet{cirtain2007} we fail to identify a separate ordinary jet travelling at non-Alfv\'enic velocities that accompanies the PMJs in our observations. \citet{esteban2019} proposed the same PMJ propagation mechanism in order to reconcile their low absolute LOS velocities and their complex LOS velocity stratifications of their PMJs relative to the high POS velocities reported by previous investigations \citep{katsukawa2007,tiwari2016}. 

The magnetic field evolution in Fig. \ref{fig:weakB} has shown that PMJs are not only an intensity feature, but also produce a measurable increase in the LOS magnetic field strength, i.e. 100 G for the PMJ in Fig. \ref{fig:composite2}. Furthermore, the magnetic field enhancement propagates at the same velocity as the PMJ's POS velocity. Given that the simulations presented by \citet{depontieu2017} and \citet{martinezsykora2017} produced Alfv\'en waves, which potentially drive the electric current that causes the heat front in their simulations, it is possible that the magnetic field evolution in Fig. \ref{fig:weakB} pertains to an Alfv\'en wave travelling along the PMJ's propagation axis. However, most of our selected PMJs have no significant Stokes $Q$, $U$, or $V$ signal except for two where the Stokes $V$ amplitude is above $3\sigma$. Therefore, better observations with access to the full magnetic field vector are necessary before the magnetic field evolution within a PMJ can be robustly interpreted in particular with regards to opacity effects.   

The more comprehensive STiC inversions carried out by \citet{esteban2019} corroborated the results of our test inversions. More importantly, the magnitude of the LOS velocities retrieved by \citet{esteban2019} and our inversions are on average of the same magnitude as the H$\alpha$ bisector LOS velocities presented by us, which further strengthens our assertion that PMJs can be detected in H$\alpha$. The inversions by \citet{esteban2019} and ours both indicate that PMJs occur near the temperature minimum above $\log(\tau)=-2$. The different $\mu$ values of our observations expresses themselves, as expected, in the form of the generally lower temperatures and magnetic field strengths retrieved by our inversions.  

The gradual brightening phase illustrated in Fig. \ref{fig:CaI} agrees with the results presented by \citet{reardon2013}, in particular, with those provided by \citet{samanta2017}. The assertion by \citet{samanta2017} that PMJs host a preceding brightening phase seen in the upper chromosphere and transition region is confirmed by our results. However, the conclusion drawn by \citet{samanta2017} that PMJs are triggered by a reconnection event in the transition region is not supported by our results.

\section{Conclusion} \label{sec:con}

In this study we present an analysis of 10 selected PMJs found in the sunspot in active region NOAA 12599, which we observed in H$\alpha$, Ca II 8542 \AA, and the Fe I 6302 \AA$ $ line pair using SST/CRISP and in Ca II K using SST/CHROMIS on the $12^{th}$ October 2016 with $\mu=0.68$.

We demonstrate for the first time  that PMJs appear as a dark feature in H$\alpha$ and are cospatial with the familiar bright, jet-like appearance in Ca II K and Ca II 8542 \AA. The PMJs' LOS velocities in H$\alpha$, on average $-1.1\pm0.6$ km/s, match the velocities we retrieve from inversions of the Fe I and Ca II 8542 \AA$ $ lines, further substantiating our assertion that PMJs are traceable in H$\alpha$. Based on the H$\alpha$ observations we are also able to extend the average length and lifetime of PMJs to $2815\pm530$ km and $163\pm25$ s, respectively. Additionally, we measured the POS velocities of our selected PMJs, which tend to give increasing velocities with increasing distance from their respective origins reaching velocities of up to $17$ km/s. Furthermore, two of our PMJs have significant Stokes $V$ signal and indicate that the PMJs exhibit an increased LOS magnetic field of up to 100 G when compared to their local pre-/post- PMJ magnetic field. The PMJs' magnetic field enhancement propagates as quickly as the PMJs' POS velocity. Finally, we present evidence that PMJs display a gradual precursory brightening, lasting for 1 minute on average, that only manifests itself in the cores of Ca II K and 8542 \AA$ $ lines.

The uncharacteristic POS velocity evolution of our PMJs, combined with their low LOS velocities, leads us to conclude that PMJs are unlikely to be ordinary jets with mass flows, but rather are manifestations of heat fronts propagating at local Alfv\'enic velocities, that dissipate their energy as they travel through the chromosphere. A future investigation should aim to obtain a greater sample of PMJs to more thoroughly ascertain their magnetic properties in particular.

\begin{acknowledgements} 
This research has been supported by the Knut and Alice Wallenberg Foundation. J.d.l.C.R. is supported by grants from the Swedish Research Council (2015-03994), the Swedish National Space Board (128/15), and the Swedish Civil Contingencies Agency (MSB). This project has received funding from the European Research Council (ERC) under the European Union's Horizon 2020 research and innovation programme (SUNMAG, grant agreement 759548). This paper is based on data acquired at the Swedish 1 m Solar Telescope, operated by the Institute for Solar Physics of Stockholm University in the Spanish Observatorio del Roque de los Muchachos of the Instituto de Astrof\'isica de Canarias. The Institute for Solar Physics is supported by a grant for research infrastructures of national importance from the Swedish Research Council (registration number 2017-00625).
\end{acknowledgements}




\bibliographystyle{apj}
\bibliography{/Users/davidbuehler/Documents/Latex/TheBib_copy}

\begin{thebibliography}{}
\expandafter\ifx\csname natexlab\endcsname\relax\def\natexlab#1{#1}\fi

\bibitem[{{Centeno}(2018)}]{centeno2018}
{Centeno}, R. 2018, \apj, 866, 89

\bibitem[{{Cirtain} {et~al.}(2007){Cirtain}, {Golub}, {Lundquist}, {van
  Ballegooijen}, {Savcheva}, {Shimojo}, {DeLuca}, {Tsuneta}, {Sakao}, {Reeves},
  {Weber}, {Kano}, {Narukage}, \& {Shibasaki}}]{cirtain2007}
{Cirtain}, J.~W., {Golub}, L., {Lundquist}, L., {et~al.} 2007, Science, 318,
  1580

\bibitem[{{Cirtain} {et~al.}(2013){Cirtain}, {Golub}, {Winebarger}, {de
  Pontieu}, {Kobayashi}, {Moore}, {Walsh}, {Korreck}, {Weber}, {McCauley},
  {Title}, {Kuzin}, \& {Deforest}}]{cirtain2013}
{Cirtain}, J.~W., {Golub}, L., {Winebarger}, A.~R., {et~al.} 2013, \nat, 493,
  501

\bibitem[{{de la Cruz Rodr{\'{\i}}guez} {et~al.}(2016){de la Cruz
  Rodr{\'{\i}}guez}, {Leenaarts}, \& {Asensio Ramos}}]{delacruz2016}
{de la Cruz Rodr{\'{\i}}guez}, J., {Leenaarts}, J., \& {Asensio Ramos}, A.
  2016, \apjl, 830, L30

\bibitem[{{de la Cruz Rodr{\'{\i}}guez} {et~al.}(2019){de la Cruz
  Rodr{\'{\i}}guez}, {Leenaarts}, {Danilovic}, \& {Uitenbroek}}]{delacruz2019}
{de la Cruz Rodr{\'{\i}}guez}, J., {Leenaarts}, J., {Danilovic}, S., \&
  {Uitenbroek}, H. 2019, \aap, 623, A74

\bibitem[{{de la Cruz Rodr{\'{\i}}guez} {et~al.}(2015){de la Cruz
  Rodr{\'{\i}}guez}, {L{\"o}fdahl}, {S{\"u}tterlin}, {Hillberg}, \& {Rouppe van
  der Voort}}]{delacruz2015}
{de la Cruz Rodr{\'{\i}}guez}, J., {L{\"o}fdahl}, M.~G., {S{\"u}tterlin}, P.,
  {Hillberg}, T., \& {Rouppe van der Voort}, L. 2015, \aap, 573, A40

\bibitem[{{De Pontieu} {et~al.}(2007){De Pontieu}, {Hansteen}, {Rouppe van der
  Voort}, {van Noort}, \& {Carlsson}}]{depontieu2007df}
{De Pontieu}, B., {Hansteen}, V.~H., {Rouppe van der Voort}, L., {van Noort},
  M., \& {Carlsson}, M. 2007, \apj, 655, 624

\bibitem[{{De Pontieu} {et~al.}(2001){De Pontieu}, {Martens}, \&
  {Hudson}}]{depontieu2001}
{De Pontieu}, B., {Martens}, P.~C.~H., \& {Hudson}, H.~S. 2001, \apj, 558, 859

\bibitem[{{De Pontieu} {et~al.}(2017){De Pontieu}, {Mart{\'{\i}}nez-Sykora}, \&
  {Chintzoglou}}]{depontieu2017}
{De Pontieu}, B., {Mart{\'{\i}}nez-Sykora}, J., \& {Chintzoglou}, G. 2017,
  \apjl, 849, L7

\bibitem[{{De Pontieu} {et~al.}(2014){De Pontieu}, {Title}, {Lemen}, {Kushner},
  {Akin}, {Allard}, {Berger}, {Boerner}, {Cheung}, {Chou}, {Drake}, {Duncan},
  {Freeland}, {Heyman}, {Hoffman}, {Hurlburt}, {Lindgren}, {Mathur}, {Rehse},
  {Sabolish}, {Seguin}, {Schrijver}, {Tarbell}, {W{\"u}lser}, {Wolfson},
  {Yanari}, {Mudge}, {Nguyen-Phuc}, {Timmons}, {van Bezooijen}, {Weingrod},
  {Brookner}, {Butcher}, {Dougherty}, {Eder}, {Knagenhjelm}, {Larsen},
  {Mansir}, {Phan}, {Boyle}, {Cheimets}, {DeLuca}, {Golub}, {Gates}, {Hertz},
  {McKillop}, {Park}, {Perry}, {Podgorski}, {Reeves}, {Saar}, {Testa}, {Tian},
  {Weber}, {Dunn}, {Eccles}, {Jaeggli}, {Kankelborg}, {Mashburn}, {Pust},
  {Springer}, {Carvalho}, {Kleint}, {Marmie}, {Mazmanian}, {Pereira}, {Sawyer},
  {Strong}, {Worden}, {Carlsson}, {Hansteen}, {Leenaarts}, {Wiesmann},
  {Aloise}, {Chu}, {Bush}, {Scherrer}, {Brekke}, {Martinez-Sykora}, {Lites},
  {McIntosh}, {Uitenbroek}, {Okamoto}, {Gummin}, {Auker}, {Jerram}, {Pool}, \&
  {Waltham}}]{depontieu2014}
{De Pontieu}, B., {Title}, A.~M., {Lemen}, J.~R., {et~al.} 2014, \solphys, 289,
  2733

\bibitem[{{Drews} \& {Rouppe van der Voort}(2017)}]{drews2017}
{Drews}, A., \& {Rouppe van der Voort}, L. 2017, \aap, 602, A80

\bibitem[{{Esteban Pozuelo} {et~al.}(2019){Esteban Pozuelo}, {de la Cruz
  Rodr{\'{\i}}guez}, {Drews}, {Rouppe van der Voort}, {Scharmer}, \&
  {Carlsson}}]{esteban2019}
{Esteban Pozuelo}, S., {de la Cruz Rodr{\'{\i}}guez}, J., {Drews}, A., {et~al.}
  2019, \apj, 870, 88

\bibitem[{{Grant} {et~al.}(2018){Grant}, {Jess}, {Zaqarashvili}, {Beck},
  {Socas-Navarro}, {Aschwanden}, {Keys}, {Christian}, {Houston}, \&
  {Hewitt}}]{grant2018}
{Grant}, S.~D.~T., {Jess}, D.~B., {Zaqarashvili}, T.~V., {et~al.} 2018, Nature
  Physics, 14, 480

\bibitem[{{Hansteen} {et~al.}(2006){Hansteen}, {De Pontieu}, {Rouppe van der
  Voort}, {van Noort}, \& {Carlsson}}]{hansteen2006}
{Hansteen}, V.~H., {De Pontieu}, B., {Rouppe van der Voort}, L., {van Noort},
  M., \& {Carlsson}, M. 2006, \apjl, 647, L73

\bibitem[{{Heggland} {et~al.}(2011){Heggland}, {Hansteen}, {De Pontieu}, \&
  {Carlsson}}]{heggland2011}
{Heggland}, L., {Hansteen}, V.~H., {De Pontieu}, B., \& {Carlsson}, M. 2011,
  \apj, 743, 142

\bibitem[{{Joshi} {et~al.}(2017){Joshi}, {Lagg}, {Hirzberger}, \&
  {Solanki}}]{joshi2017}
{Joshi}, J., {Lagg}, A., {Hirzberger}, J., \& {Solanki}, S.~K. 2017, \aap, 604,
  A98

\bibitem[{{Joshi} {et~al.}(2011){Joshi}, {Pietarila}, {Hirzberger}, {Solanki},
  {Aznar Cuadrado}, \& {Merenda}}]{joshi2011}
{Joshi}, J., {Pietarila}, A., {Hirzberger}, J., {et~al.} 2011, \apjl, 734, L18

\bibitem[{{Jur{\v c}{\'a}k} \& {Katsukawa}(2008)}]{jurcak2008}
{Jur{\v c}{\'a}k}, J., \& {Katsukawa}, Y. 2008, \aap, 488, L33

\bibitem[{{Katsukawa} \& {Jur{\v c}{\'a}k}(2010)}]{katsukawa2010}
{Katsukawa}, Y., \& {Jur{\v c}{\'a}k}, J. 2010, \aap, 524, A20

\bibitem[{{Katsukawa} {et~al.}(2007){Katsukawa}, {Berger}, {Ichimoto}, {Lites},
  {Nagata}, {Shimizu}, {Shine}, {Suematsu}, {Tarbell}, {Title}, \&
  {Tsuneta}}]{katsukawa2007}
{Katsukawa}, Y., {Berger}, T.~E., {Ichimoto}, K., {et~al.} 2007, Science, 318,
  1594

\bibitem[{{Kosugi} {et~al.}(2007){Kosugi}, {Matsuzaki}, {Sakao}, {Shimizu},
  {Sone}, {Tachikawa}, {Hashimoto}, {Minesugi}, {Ohnishi}, {Yamada}, {Tsuneta},
  {Hara}, {Ichimoto}, {Suematsu}, {Shimojo}, {Watanabe}, {Shimada}, {Davis},
  {Hill}, {Owens}, {Title}, {Culhane}, {Harra}, {Doschek}, \&
  {Golub}}]{kosugi2007}
{Kosugi}, T., {Matsuzaki}, K., {Sakao}, T., {et~al.} 2007, \solphys, 243, 3

\bibitem[{{Landi Degl'Innocenti} \& {Landolfi}(2004)}]{landi2004}
{Landi Degl'Innocenti}, E., \& {Landolfi}, M., eds. 2004, Astrophysics and
  Space Science Library, Vol. 307, {Polarization in Spectral Lines}

\bibitem[{{Lemen} {et~al.}(2012){Lemen}, {Title}, {Akin}, {Boerner}, {Chou},
  {Drake}, {Duncan}, {Edwards}, {Friedlaender}, {Heyman}, {Hurlburt}, {Katz},
  {Kushner}, {Levay}, {Lindgren}, {Mathur}, {McFeaters}, {Mitchell}, {Rehse},
  {Schrijver}, {Springer}, {Stern}, {Tarbell}, {Wuelser}, {Wolfson}, {Yanari},
  {Bookbinder}, {Cheimets}, {Caldwell}, {Deluca}, {Gates}, {Golub}, {Park},
  {Podgorski}, {Bush}, {Scherrer}, {Gummin}, {Smith}, {Auker}, {Jerram},
  {Pool}, {Soufli}, {Windt}, {Beardsley}, {Clapp}, {Lang}, \&
  {Waltham}}]{lemen2012}
{Lemen}, J.~R., {Title}, A.~M., {Akin}, D.~J., {et~al.} 2012, \solphys, 275, 17

\bibitem[{{Lites} {et~al.}(1993){Lites}, {Rutten}, \& {Kalkofen}}]{lites1993}
{Lites}, B.~W., {Rutten}, R.~J., \& {Kalkofen}, W. 1993, \apj, 414, 345

\bibitem[{{L{\"o}fdahl} {et~al.}(2018){L{\"o}fdahl}, {Hillberg}, {de la Cruz
  Rodriguez}, {Vissers}, {Scharmer}, {Hagfors Haugan}, \&
  {Fredvik}}]{lofdahl2018}
{L{\"o}fdahl}, M.~G., {Hillberg}, T., {de la Cruz Rodriguez}, J., {et~al.}
  2018, ArXiv e-prints, arXiv:1804.03030

\bibitem[{{Magara}(2010)}]{magara2010}
{Magara}, T. 2010, \apjl, 715, L40

\bibitem[{{Mart{\'{\i}}nez-Sykora} {et~al.}(2017){Mart{\'{\i}}nez-Sykora}, {De
  Pontieu}, {Hansteen}, {Rouppe van der Voort}, {Carlsson}, \&
  {Pereira}}]{martinezsykora2017}
{Mart{\'{\i}}nez-Sykora}, J., {De Pontieu}, B., {Hansteen}, V.~H., {et~al.}
  2017, Science, 356, 1269

\bibitem[{{Narang} {et~al.}(2016){Narang}, {Arbacher}, {Tian}, {Banerjee},
  {Cranmer}, {DeLuca}, \& {McKillop}}]{narang2016}
{Narang}, N., {Arbacher}, R.~T., {Tian}, H., {et~al.} 2016, \solphys, 291, 1129

\bibitem[{{Reardon} {et~al.}(2013){Reardon}, {Tritschler}, \&
  {Katsukawa}}]{reardon2013}
{Reardon}, K., {Tritschler}, A., \& {Katsukawa}, Y. 2013, \apj, 779, 143

\bibitem[{{Robustini} {et~al.}(2016){Robustini}, {Leenaarts}, {de la Cruz
  Rodriguez}, \& {Rouppe van der Voort}}]{robustini2016}
{Robustini}, C., {Leenaarts}, J., {de la Cruz Rodriguez}, J., \& {Rouppe van
  der Voort}, L. 2016, \aap, 590, A57

\bibitem[{{Rouppe van der Voort} \& {de la Cruz
  Rodr{\'{\i}}guez}(2013)}]{vandervoort2013}
{Rouppe van der Voort}, L., \& {de la Cruz Rodr{\'{\i}}guez}, J. 2013, \apj,
  776, 56

\bibitem[{{Rouppe van der Voort} {et~al.}(2015){Rouppe van der Voort}, {De
  Pontieu}, {Pereira}, {Carlsson}, \& {Hansteen}}]{vandervoort2015}
{Rouppe van der Voort}, L., {De Pontieu}, B., {Pereira}, T.~M.~D., {Carlsson},
  M., \& {Hansteen}, V. 2015, \apjl, 799, L3

\bibitem[{{Rouppe van der Voort} {et~al.}(2009){Rouppe van der Voort},
  {Leenaarts}, {de Pontieu}, {Carlsson}, \& {Vissers}}]{vandervoort2009}
{Rouppe van der Voort}, L., {Leenaarts}, J., {de Pontieu}, B., {Carlsson}, M.,
  \& {Vissers}, G. 2009, \apj, 705, 272

\bibitem[{{Ruiz Cobo} \& {Asensio Ramos}(2013)}]{ruizcobo2013}
{Ruiz Cobo}, B., \& {Asensio Ramos}, A. 2013, \aap, 549, L4

\bibitem[{{Ryutova} {et~al.}(2008){Ryutova}, {Berger}, {Frank}, \&
  {Title}}]{ryutova2008}
{Ryutova}, M., {Berger}, T., {Frank}, Z., \& {Title}, A. 2008, \apj, 686, 1404

\bibitem[{{Sakai} \& {Smith}(2008)}]{sakai2008}
{Sakai}, J.~I., \& {Smith}, P.~D. 2008, \apjl, 687, L127

\bibitem[{{Samanta} {et~al.}(2017){Samanta}, {Tian}, {Banerjee}, \&
  {Schanche}}]{samanta2017}
{Samanta}, T., {Tian}, H., {Banerjee}, D., \& {Schanche}, N. 2017, \apjl, 835,
  L19

\bibitem[{{Scharmer} {et~al.}(2003){Scharmer}, {Bjelksjo}, {Korhonen},
  {Lindberg}, \& {Petterson}}]{scharmer2003}
{Scharmer}, G.~B., {Bjelksjo}, K., {Korhonen}, T.~K., {Lindberg}, B., \&
  {Petterson}, B. 2003, in Society of Photo-Optical Instrumentation Engineers
  (SPIE) Conference Series, Vol. 4853, Innovative Telescopes and
  Instrumentation for Solar Astrophysics, ed. S.~L. {Keil} \& S.~V. {Avakyan},
  341--350

\bibitem[{{Scharmer} {et~al.}(2013){Scharmer}, {de la Cruz Rodriguez},
  {S{\"u}tterlin}, \& {Henriques}}]{scharmer2013}
{Scharmer}, G.~B., {de la Cruz Rodriguez}, J., {S{\"u}tterlin}, P., \&
  {Henriques}, V.~M.~J. 2013, \aap, 553, A63

\bibitem[{{Scharmer} {et~al.}(2011){Scharmer}, {Henriques}, {Kiselman}, \& {de
  la Cruz Rodr{\'{\i}}guez}}]{scharmer2011}
{Scharmer}, G.~B., {Henriques}, V.~M.~J., {Kiselman}, D., \& {de la Cruz
  Rodr{\'{\i}}guez}, J. 2011, Science, 333, 316

\bibitem[{{Scharmer} {et~al.}(2008){Scharmer}, {Narayan}, {Hillberg}, {de la
  Cruz Rodriguez}, {L{\"o}fdahl}, {Kiselman}, {S{\"u}tterlin}, {van Noort}, \&
  {Lagg}}]{scharmer2008}
{Scharmer}, G.~B., {Narayan}, G., {Hillberg}, T., {et~al.} 2008, \apjl, 689,
  L69

\bibitem[{{Sekse} {et~al.}(2012){Sekse}, {Rouppe van der Voort}, \& {De
  Pontieu}}]{sekse2012}
{Sekse}, D.~H., {Rouppe van der Voort}, L., \& {De Pontieu}, B. 2012, \apj,
  752, 108

\bibitem[{{Sekse} {et~al.}(2013){Sekse}, {Rouppe van der Voort}, {De Pontieu},
  \& {Scullion}}]{sekse2013}
{Sekse}, D.~H., {Rouppe van der Voort}, L., {De Pontieu}, B., \& {Scullion}, E.
  2013, \apj, 769, 44

\bibitem[{{Solanki}(2003)}]{solanki2003rev}
{Solanki}, S.~K. 2003, \aapr, 11, 153

\bibitem[{{Tian} {et~al.}(2014){Tian}, {DeLuca}, {Cranmer}, {De Pontieu},
  {Peter}, {Mart{\'{\i}}nez-Sykora}, {Golub}, {McKillop}, {Reeves}, {Miralles},
  {McCauley}, {Saar}, {Testa}, {Weber}, {Murphy}, {Lemen}, {Title}, {Boerner},
  {Hurlburt}, {Tarbell}, {Wuelser}, {Kleint}, {Kankelborg}, {Jaeggli},
  {Carlsson}, {Hansteen}, \& {McIntosh}}]{tian2014}
{Tian}, H., {DeLuca}, E.~E., {Cranmer}, S.~R., {et~al.} 2014, Science, 346,
  1255711

\bibitem[{{Tiwari} {et~al.}(2018){Tiwari}, {Moore}, {De Pontieu}, {Tarbell},
  {Panesar}, {Winebarger}, \& {Sterling}}]{tiwari2018}
{Tiwari}, S.~K., {Moore}, R.~L., {De Pontieu}, B., {et~al.} 2018, ArXiv
  e-prints, arXiv:1811.09554

\bibitem[{{Tiwari} {et~al.}(2016){Tiwari}, {Moore}, {Winebarger}, \&
  {Alpert}}]{tiwari2016}
{Tiwari}, S.~K., {Moore}, R.~L., {Winebarger}, A.~R., \& {Alpert}, S.~E. 2016,
  \apj, 816, 92

\bibitem[{{Tiwari} {et~al.}(2013){Tiwari}, {van Noort}, {Lagg}, \&
  {Solanki}}]{tiwari2013}
{Tiwari}, S.~K., {van Noort}, M., {Lagg}, A., \& {Solanki}, S.~K. 2013, \aap,
  557, A25

\bibitem[{{Uitenbroek}(2001)}]{uitenbroek2001}
{Uitenbroek}, H. 2001, \apj, 557, 389

\bibitem[{{van Noort} {et~al.}(2005){van Noort}, {Rouppe van der Voort}, \&
  {L{\"o}fdahl}}]{vannoort2005}
{van Noort}, M., {Rouppe van der Voort}, L., \& {L{\"o}fdahl}, M.~G. 2005,
  \solphys, 228, 191

\bibitem[{{van Noort} \& {Rouppe van der Voort}(2008)}]{vannoort2008}
{van Noort}, M.~J., \& {Rouppe van der Voort}, L.~H.~M. 2008, \aap, 489, 429

\bibitem[{{Vissers} \& {Rouppe van der Voort}(2012)}]{vissers2012}
{Vissers}, G., \& {Rouppe van der Voort}, L. 2012, \apj, 750, 22

\bibitem[{{Vissers} {et~al.}(2015){Vissers}, {Rouppe van der Voort}, \&
  {Carlsson}}]{vissers2015}
{Vissers}, G.~J.~M., {Rouppe van der Voort}, L.~H.~M., \& {Carlsson}, M. 2015,
  \apjl, 811, L33

\bibitem[{{Zakharov} {et~al.}(2008){Zakharov}, {Hirzberger}, {Riethm{\"u}ller},
  {Solanki}, \& {Kobel}}]{zakharov2008}
{Zakharov}, V., {Hirzberger}, J., {Riethm{\"u}ller}, T.~L., {Solanki}, S.~K.,
  \& {Kobel}, P. 2008, \aap, 488, L17

\end{thebibliography}

\begin{appendix}

\begin{deluxetable}{cccccc}
\tablecaption{Properties of PMJs \label{tab:properties}}
\tablecolumns{6}
\tablenum{1}
\tablewidth{0pt}
\tablehead{
\colhead{Length} &
\colhead{Lifetime} &
\colhead{Width} & 
\colhead{LOS velocity\tablenotemark{a}} & 
\colhead{Length\tablenotemark{b}} &
\colhead{Lifetime\tablenotemark{b}} \\
\colhead{[km]} & 
\colhead{[s]} &
\colhead{[km]} & 
\colhead{[km/s]} & 
\colhead{[km]} &
\colhead{[s]}
}
\startdata
3547 & 234 & 308 & -2.2 & 547 & 91\\
3004 & 156 & 194 & -1.3 & 678 & 52\\
2525 & 156 & 363 & -0.8 & 492 & 52\\
2402 & 120 & 291 & -1.0 & 1102 & 91\\
2465 & 156 & 205 & -1.0 & 834 & 78\\
3161 & 180 & 176 & -2.0 & 701 & 65\\
3537 & 180 & 211 & -0.8 & 626 & 52\\
2393 & 156 & 185 & -0.5 & 442 & 65\\
2000 & 156 & 209 & -0.2 & 842 & 65\\
1561 & 156 & 220 & -1.0 & 676 & 91\\
\enddata
\tablenotetext{a}{mean LOS velocity from H$\alpha$ bisectors}
\tablenotetext{b}{values obtained using only the wings of Ca II 8542 \AA$ $ and Ca II K (e.g. bottom row in Fig. \ref{fig:composite2})}
\tablecomments{ The distances, lifetimes, and velocities carry uncertainties of $\pm60$ km, $\pm13$ s, and $\pm0.2$ km/s, respectively.}
\end{deluxetable}

\begin{figure*}
\includegraphics[width=18cm]{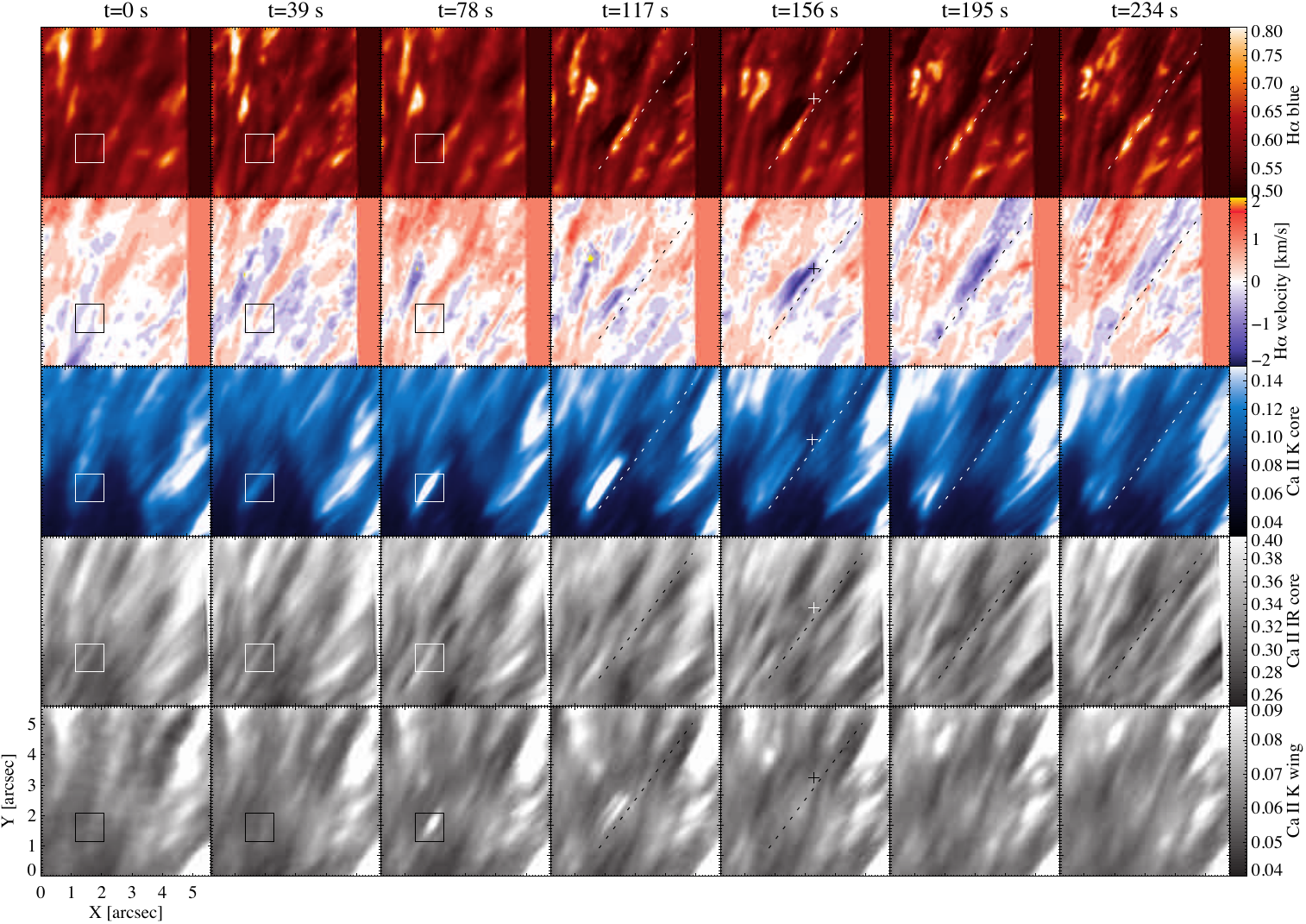}
\caption{Sequence of H$\alpha$ -450 m\AA, H$\alpha$ bisector LOS velocities at -450 m\AA, Ca II K -60 m\AA, Ca II 8542 -85 m\AA, and Ca II K -300 m\AA$ $ images. The squares enclose the origin of the PMJ that occurs at $t=78$ s, and the plus symbols have been added for reference. The dotted lines are in parallel with the PMJ's propagation axis and the plus symbols serve as a common point of reference. The location of the PMJ is marked by the diamond symbol in Fig. \ref{fig:composite1}. \label{fig:compositeappen}}
\end{figure*}

\end{appendix}



\end{document}